\documentclass[useAMS]{mn2e} 
 
\usepackage{graphicx} 
\usepackage{txfonts} 
\newcommand\aj{AJ} 
\newcommand\apj{ApJ} 
\newcommand\apjs{ApJS}       
 
\newcommand\aap{A\&A} 
\newcommand\mnras{MNRAS} 
\newcommand\apjl{ApJ} 
\newcommand\pasp{PASP} 
\newcommand\nat{Nature}

\title[]{Four stellar populations and extreme helium variation in the massive outer-halo globular cluster NGC\,2419}
\author[ M.\,Zennaro et al.] 
       {M.\,Zennaro$^{1}$,
         A.\,P.\,Milone$^{1}$,
         A.\,F.\,Marino$^{1,2}$,
         G.\,Cordoni$^{1}$,
         E.\,P.\,Lagioia$^{1}$,
         M.\,Tailo$^{1}$
\\ 
$^{1}$Dipartimento di Fisica e Astronomia ``Galileo Galilei'', Univ. di Padova, Vicolo dell'Osservatorio 3, Padova, IT-35122\\
$^{2}$Centro di Ateneo di Studi e Attivita’ Spaziali “Giuseppe Colombo” - CISAS, Via Venezia 15, Padova, IT-35131\\
       } 
\begin{document} 
\pagerange{\pageref{firstpage}--\pageref{lastpage}} \pubyear{2017} 
 
\maketitle 
\label{firstpage} 
 
\begin{abstract}
   Recent work revealed that both the helium variation within globular clusters (GCs) and the relative numbers of first and second-generation stars (1G, 2G) depend on the mass of the host cluster. Precise determination of the internal helium variations and of the fraction of 1G stars are crucial constraints to the formation scenarios of multiple populations (MPs).

 We exploit multi-band  {\it Hubble Space Telescope} photometry to investigate MPs in NGC\,2419, which is one of the most-massive and distant GCs of the Galaxy, almost isolated from its tidal influence. We find that the 1G hosts the $\sim$37\% of the analyzed stars, and identified three populations of 2G stars, namely G$_{\rm A}$, 2G$_{\rm B}$, and 2G$_{\rm C}$, which comprise the $\sim$20\%, $\sim$31\% and $\sim$12\% of stars, respectively.
  
  We compare the observed colors of these four populations with the colors derived from appropriate synthetic spectra to infer the relative helium abundances. We find that 2G$_{\rm A}$, 2G$_{\rm B}$, and 2G$_{\rm C}$ stars are enhanced in helium mass fraction by $\delta$Y$\sim$0.01, 0.06, and 0.19 with respect to 1G stars that have primordial helium (Y=0.246).

The high He enrichment of 2G$_{\rm C}$ stars is hardly reconcilable with most of the current scenarios for MPs.
Furthermore, the relatively larger fraction of 1G stars ($\sim$37\%) compared to other massive GCs is noticeable. By exploiting literature results, we find that the fractions of 1G stars of GCs with large perigalactic distance are typically higher than in the other GCs with similar masses. This suggests that NGC\,2419, similarly to other distant GCs, lost a lower fraction of 1G stars.

\end{abstract}
\begin{keywords} 
\end{keywords} 
 
\section{Introduction}\label{sec:intro} 
Nearly all GCs are composed of two main groups of first- and second-generation stars (1G, 2G) with different chemical compositions whose origin is still not satisfactory understood (Milone et al.\,2017).
 According to many scenarios, 2G stars formed in the cluster center out of the material polluted by more-massive 1G stars (e.g.\,Ventura et al.\,2001; Decressin et al.\,2007; D'Ercole et al.\,2008, 2010). In these scenarios, GCs would lose the majority of their 1G, thus providing a significant contribution to the assembly of the Galaxy (e.g.\,D'Ercole et al.\,2010; D'Antona et al.\,2016). As an alternative, GCs would host a single stellar generation and stars with different chemical composition are the product of exotic phenomena that occur in the unique environment of proto GCs (e.g.\,De Mink et al.\,2009; Bastian et al.\,2013; Gieles et al.\,2018).

NGC\,2419 is one of the most-massive ($\mathcal{M}= 9 \cdot 10^{5} \mathcal{M_{\odot}}$, McLaughlin \& van der Marel\,2005) and metal-poor ([Fe/H]=$-$2.09, Mucciarelli et al.\,2012) Galactic GCs. The large distance from the Galactic center ($d \sim 87.5$ kpc, Di Criscienzo et al.\,2011a) makes it almost isolated from the tidal influence of the Milky Way. Moreover, since its half-light relaxation time exceeds the Hubble time (Harris 1996), it would retain fossil information on the properties of multiple populations at the formation. 
The possibility that NGC\,2419 evolved in isolation, together with its extreme mass and metallicity, makes this cluster an ideal target to constrain the formation scenarios of multiple populations.

 Recent work based on multi-band photometry of 58 Galactic GCs have investigated the relation between multiple populations and foundamental parameters of the host clusters and revealed that the complexity of multiple populations increases with cluster mass.
 In particular, the maximum internal helium variation, which ranges from less than 0.01 to more than $\sim$0.12 in helium mass fraction, correlates with the mass of the host cluster, whereas the fraction of 1G stars with respect to the total number of cluster stars  varies between $\sim 8\%$ and 67\% and anti-correlates with cluster mass (Milone 2015; Milone et al.\,2017, 2018).

In the context of the multiple-generation scenarios (D'Ercole et al.\,2010; D'Antona et al.\,2016 and references therein), we would expect that a cluster that formed and evolved in isolation has retained its initial mass and the fraction of 1G stars.  As a consequence, NGC\,2419 would exhibit similar helium spread as the other GCs with similar mass but a significantly larger frequency of 1G stars that would include up to $\sim$90\% of the cluster stars (e.g.\,Di Criscienzo et al.\,2011b). 

 Multiple stellar populations in NGC\,2419 have been widely investigated both spectroscopically and photometrically.
 Based on high-precision photometry obtained from Wide Field Channel of the Advanced Camera for Survey (WFC/ACS) on the {\it Hubble Space Telescope} ({\it HST}), Di Criscienzo et al.\,(2011b) and Lee et al.\,(2013) found that the base of the red-giant branch (RGB) of NGC\,2419 exhibits a wide $m_{\rm F475W}-m_{\rm F814W}$ color broadening that is consistent with two stellar populations with an extreme helium difference of $\Delta$ Y $\sim 0.17-0.19$.
 
 Spectroscopy revealed large star-to-star variation in magnesium and potassium, at variance with most GCs that have homogeneous [K/Fe] (Cohen et al\,2011, 2012; Mucciarelli et al.\,2012). Stars with different abundances of Mg and K exhibit different colors in the $V$ vs.\,$u-V$ CMD (Beccari et al.\,2013), in close analogy with what is observed in nearly all the GCs where stars with different light-element abundance populate distinct sequences in CMDs made with ultraviolet filters (Marino et al.\,2008, Yong et al.\,2008). 

 In this paper we further investigate multiple stellar populations in NGC\,2419 by extending to this cluster the same methods used by Milone et al.\,(2017, 2018) to identify and characterize multiple populations of 58 GCs. The paper is organized as follows.
Section~\ref{sec:data} describes the dataset and the data reduction. In Sect.~\ref{sec:cmd} we present various photometric diagrams that we use to identify stellar populations in NGC\,2419 and to derive the fraction of stars in each population. 
 The chemical composition of the stellar populations is investigated in Sect.~\ref{sec:He}.  Results are discussed in Sect.~\ref{sec:discussion} where we also provide a summary of the paper. 
 
\section{Data and data analysis} \label{sec:data}
To investigate multiple stellar populations in NGC\,2419 we used archive images collected through 14 filters of the Ultraviolet and Visual Channel of the Wide Field Camera 3 (UVIS/WFC3) on board of {\it HST}. These data are collected as part of the GO-11903 program (PI J.\,Kalirai) with the main purpose of improving the UVIS photometric zero points and as part of GO\,15078 (PI S.\,Larsen), which is a project focused on the dynamics of multiple populations in NGC\,2419.
The main properties of the dataset are summarized in Table~\ref{tab:data}.

Photometry and astrometry have been obtained with the computer program {\it Kitchen Sink 2} developed by Jay Anderson, which is similar to the software described by Anderson et al.\,(2008) to reduce images taken with the Wide Field Channel of the Advanced Camera for Survey but is optimized to work with images collected with various detectors of {\it HST}, including UVIS/WFC3. 

Shortly, the software performs the fitting of appropriate point-spread functions (PSFs) to all the observed sources and follows two distinct methods to measure stars with different luminosities. The magnitudes and positions of bright stars are measured in each exposure independently, and then averaged. To measure faint stars the software combines information from all the images that are placed into a common distortion-free reference frame. We used the solution provided by Bellini et al.\,(2009, 2011) to correct the geometric distortion of the UVIS/WFC3 images. We refer to papers by Sabbi et al.\,(2016) and Bellini et al.\,(2017) for details on {\it Kitchen Sink 2}.
 Photometry has been calibrated to the Vega system as in Bedin et al.\,(2005) by using the updated zero points of the UVIS/WFC3 filters provided by the STScI webpage.

 The software by Anderson and collaborators provides various diagnostics of the photometric and astrometric quality that we used to select a sample of relatively-isolated stars that are well fitted by the PSF and have small random mean scatter (rms) in magnitudes and positions. To do this, we applied the procedure by Milone et al.\,(2009) and Bedin et al.\,(2009). Finally, we corrected the magnitudes from the small variations of the photometric zero point across the field of view as in Milone et al.\,(2012).
 
\subsection{Artificial stars}
We performed artificial-star (AS) experiments  to infer the photometric uncertainties and to simulate the photometric diagrams by extending to NGC\,2419 the procedure by Anderson et al.\,(2008).

In a nutshell, we first generated a list of coordinates and magnitudes of 100,000 stars. These stars have similar spatial distribution along the field of view as cluster stars and instrumental magnitudes, $-2.5$log$_{10}$(flux), ranging from $-13.8$ to $-4.0$ in the F814W band. The other magnitudes are derived from the corresponding fiducial lines of RGB, sub-giant branch, and main-sequence (MS) stars that we derived from the observed CMDs. 

ASs are reduced by adopting exactly the same procedure used for real stars. The  {\it Kitchen Sink 2} computer program derives for ASs the same diagnostics of the photometric and astrometric quality calculated for real stars. We included in our investigation only the sample of relatively-isolated ASs that are well fitted by the PSF and have small random mean scatter in magnitudes and positions and that are selected by using the same criteria that we adopted for real stars.

\section{Multiple populations in NGC\,2419}\label{sec:cmd}
\begin{centering} 
 \begin{figure*} 
  \includegraphics[width=8.5cm]{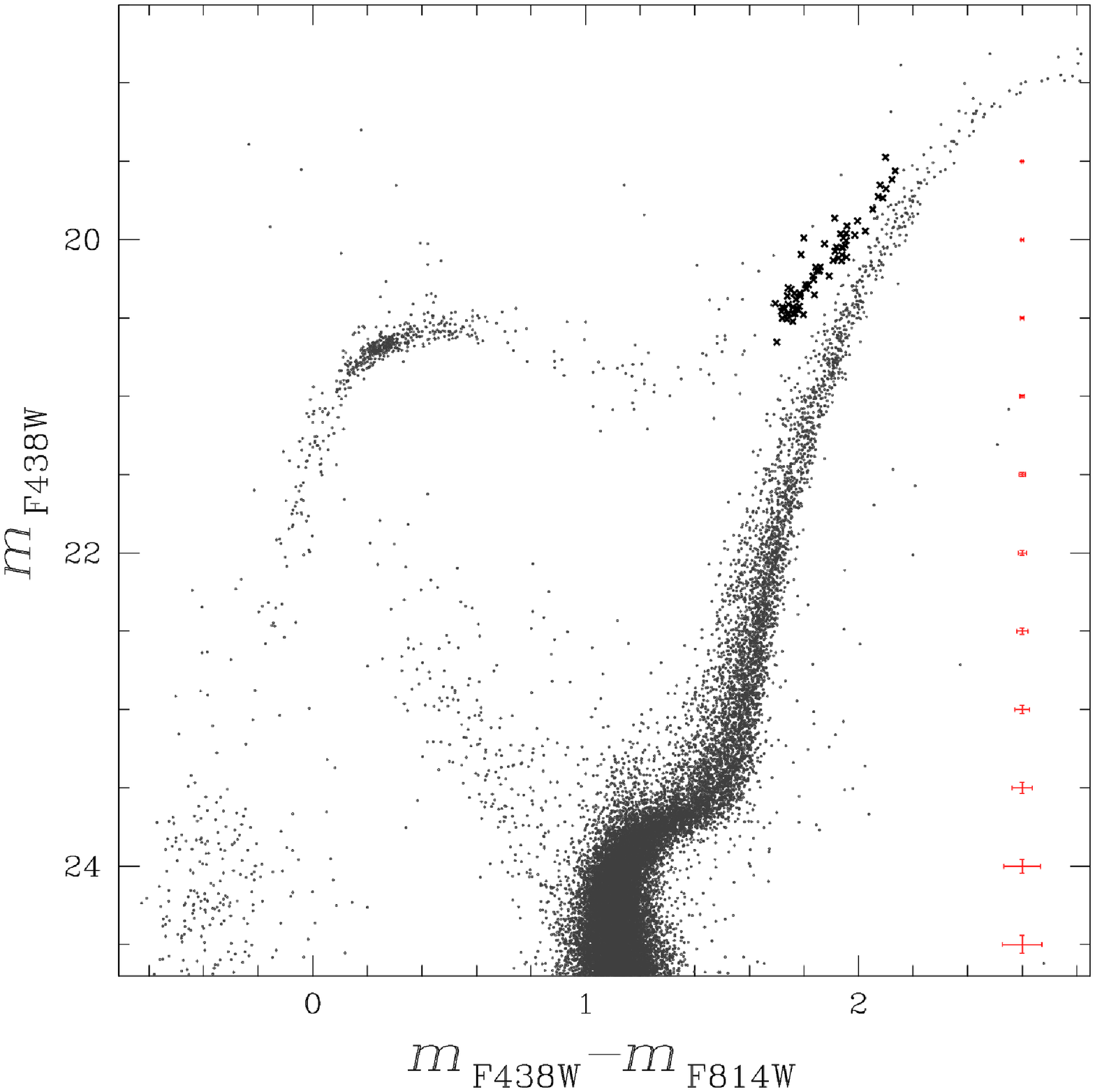}
  \includegraphics[width=8.5cm]{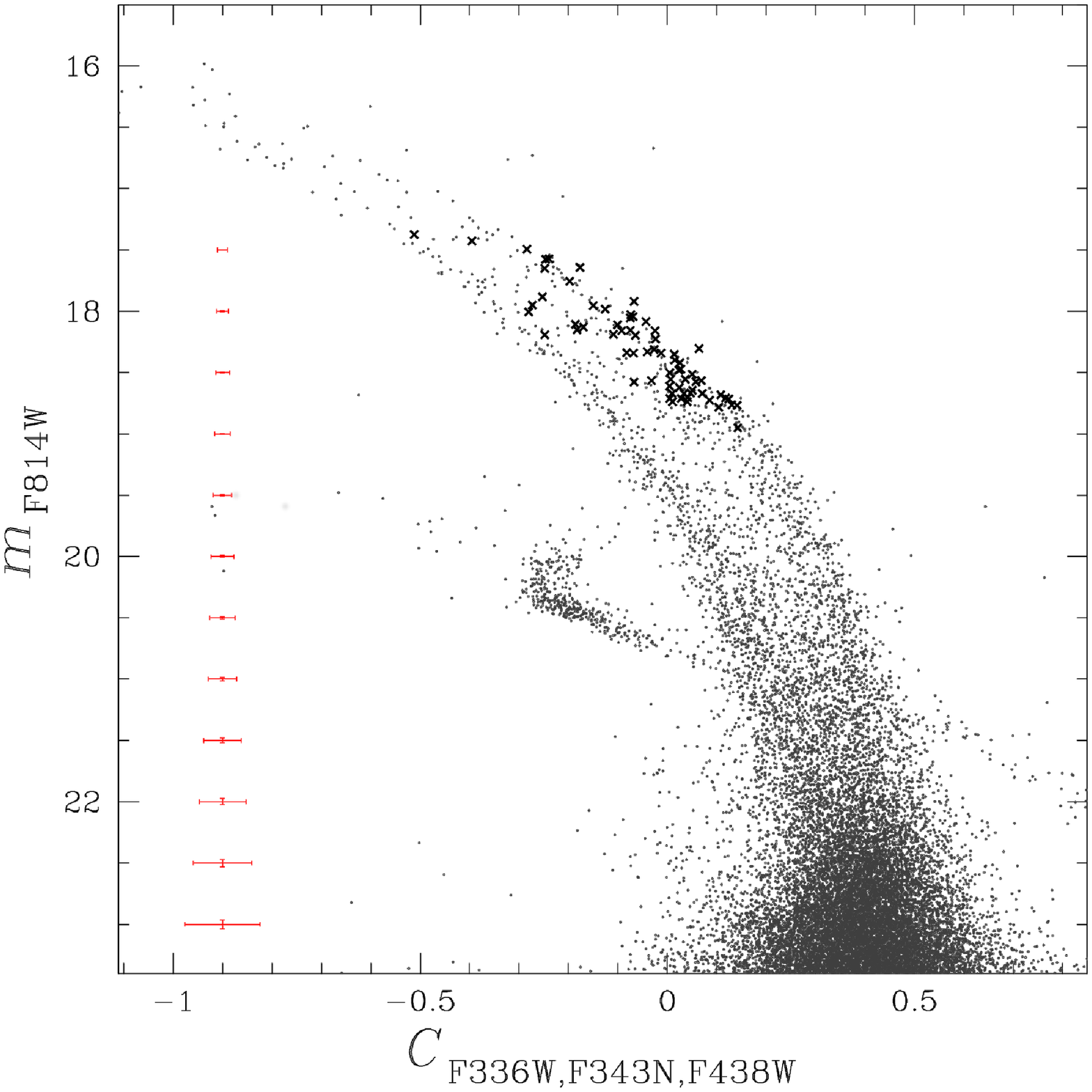}
  \caption{
    $m_{\rm F438W}$ vs.\,$m_{\rm F438W}-m_{\rm F814W}$ CMD of NGC\,2419 stars (left panel) and $m_{\rm F814W}$ against $C_{\rm F336W,F343N,F438W}$ pseudo-CMD (right panels). The typical photometric uncertainties for stars with different luminosities are indicated in each panel. Both diagrams highlight multiple stellar populations along the RGB of NGC\,2419. The AGB stars are marked with black crosses. See text for details.} 
  \label{fig:cmds} 
 \end{figure*} 
 \end{centering}

We show in Fig.~\ref{fig:cmds} two diagrams that highlight different properties of stellar populations.  The left panel of Fig.~\ref{fig:cmds} reveals that the $m_{\rm F438W}$ vs.\,$m_{\rm F438W}-m_{\rm F814W}$ CMD of NGC\,2419 is not consistent with a simple population.
Our conclusion is supported by the presence of a tail of stars with bluer colors than the bulk of RGB stars and by the fact that the color width of the RGB is much larger than what we expect from photometric uncertainties alone, which are indicated by the error bars plotted on the right of the CMD.
The fact that $m_{\rm F438W}-m_{\rm F814W}$ is an efficient color to identify RGB stars with the same luminosity but different effective temperature suggests that NGC\,2419 hosts stellar populations with extreme helium abundance as previously noticed by  Di Criscienzo et al.\,(2011b) and Lee et al.\,(2013).

In the right panel of Fig.~\ref{fig:cmds} we plotted $m_{\rm F814W}$ against $C_{\rm F336W,F343N,F438W}$=($m_{\rm F336W}-m_{\rm F343N}$)$-$($m_{\rm F343N}-m_{\rm F438W}$), which is a pseudo-color sensitive to the abundances of C and N, mostly through the NH and CN bands. The RGB is clearly split into a red and blue sequence, which include approximately 35\% and 45\% of RGB stars, respectively, plus a population of stars located between the two main RGBs that comprises about 20\% of stars.

The black crosses overimposed on both photometric diagrams of Fig.~\ref{fig:cmds} mark the asymptotic-giant branch (AGB) stars  that we selected from the left-panel CMD. Although the $C_{\rm F336W,F343N,F438W}$ broadening of AGB stars is larger than the broadening expected from observational errors, these AGB stars span a smaller range of $C_{\rm F336W,F343N,F438W}$ than RGB stars with the same luminosity. This fact indicates that, although the AGB of NGC\,2419 is not consistent with a simple population, those 2G stars with extreme chemical composition avoid the AGB phase in close analogy with what is observed in NGC\,6752, NGC\,6266 and NGC\,2808 (Campbell et al.\,2013; Lapenna et al.\,2016; Wang et al.\,2016; Marino et al.\,2017). 

\begin{table*}
  \caption{Description of the UVIS/WFC3 images of NGC\,2419 used in this paper.}

\begin{tabular}{c c c c c}
\hline \hline
    FILTER  & DATE & N$\times$EXPTIME & PROGRAM & PI \\
    \hline
     F225W & May 15 2010 & 750s & 11903 & J. Kalirai \\
     F275W & May 15 2010 & 400s & 11903 & J. Kalirai \\
     F300X & May 15 2010 & 467s & 11903 & J. Kalirai  \\
     F336W & Apr 26 2018 & $2\times1392$s$+4\times1448$s & 15078 & S. Larsen \\
     F343N & Apr 28 2018 - May 01 2018 & $4\times1392$s$+8\times1448$s & 15078 & S. Larsen \\
     F390W & May 15 2010 & 300s & 11903 & J. Kalirai \\
     F438W & May 15 2010 & $2\times 725$s & 11903 & J. Kalirai \\
     F475X & May 15 2010 & 275s & 11903 & J. Kalirai \\
     F475W & May 15 2010 & 465s & 11903 & J. Kalirai \\
     F555W & May 15 2010 & $2\times580$s &11903 & J. Kalirai \\
     F606W & May 15 2010 & $2\times400$s & 11903 & J. Kalirai \\
     F625W & May 15 2010 & 600s  & 11903 & J. Kalirai \\
     F775W & May 15 2010 & $2\times750$s & 11903 & J. Kalirai \\
     F814W & May 15 2010 & $2\times650$s & 11903 & J. Kalirai \\
     \hline\hline
\end{tabular}
  \label{tab:data}
 \end{table*}

\subsection{The chromosome map of NGC\,2419}
  The chromosome map (ChM) is a powerful tool developed to identify and characterize multiple stellar populations in GCs (Milone et al.\,2015). 
  It consists in a pseudo two-color diagram of MS, RGB or AGB stars derived from photometry in different filters that are sensitive to the specific chemical composition of the distinct populations (e.g.\,Marino et al.\,2017; Milone et al.\,2017a). The most-widely filters used to construct the ChM are F275W, F336W, F438W, and F814W of WFC3/UVIS but other optical and near-infrared bands, like F606W, F814W, F110W and F160W have been also used to derive the ChM of low-mass MS stars (Milone et al.\,2017b).

  Milone et al.\,(2017a, 2018) constructed the ChMs for RGB stars in 58 Galactic GCs by plotting the pseudo-color $C_{\rm F275W,F336W,F438W}$, which is mostly sensitive to the nitrogen abundance of the stellar populations, as a function of $m_{\rm F275W}-m_{\rm F814W}$, which is very sensitive to helium. However, the ChM is not a simple two-color diagram because the RGB is verticalized in both dimensions (see Milone et al.\,2015, 2017a for details).
  
  Unfortunately, the available F275W photometry of NGC\,2419 is obtained from a single exposure of 400s only and is too shallow to properly identify multiple populations along the RGB of this distant cluster. As a consequence, we build a ChM by using photometric bands that are different from those used by Milone and collaborators, but are sensitive to helium and nitrogen variations.
  Specifically, we combined the information from the photometric diagrams plotted in Fig.~\ref{fig:cmds} and exploited the verticalized (F438W$-$F814W) color, $\Delta_{\rm F438W,F814W}$, that is sensitive to helium, and the verticalized (F336W$-$F343N)$-$(F343N$-$F438W) pseudo-color, $\Delta_{C \rm F336W,F343N,F438W}$, which is an efficient tool to identify stellar populations with different nitrogen abundance.
  
\begin{centering} 
 \begin{figure*} 
  \includegraphics[width=13cm]{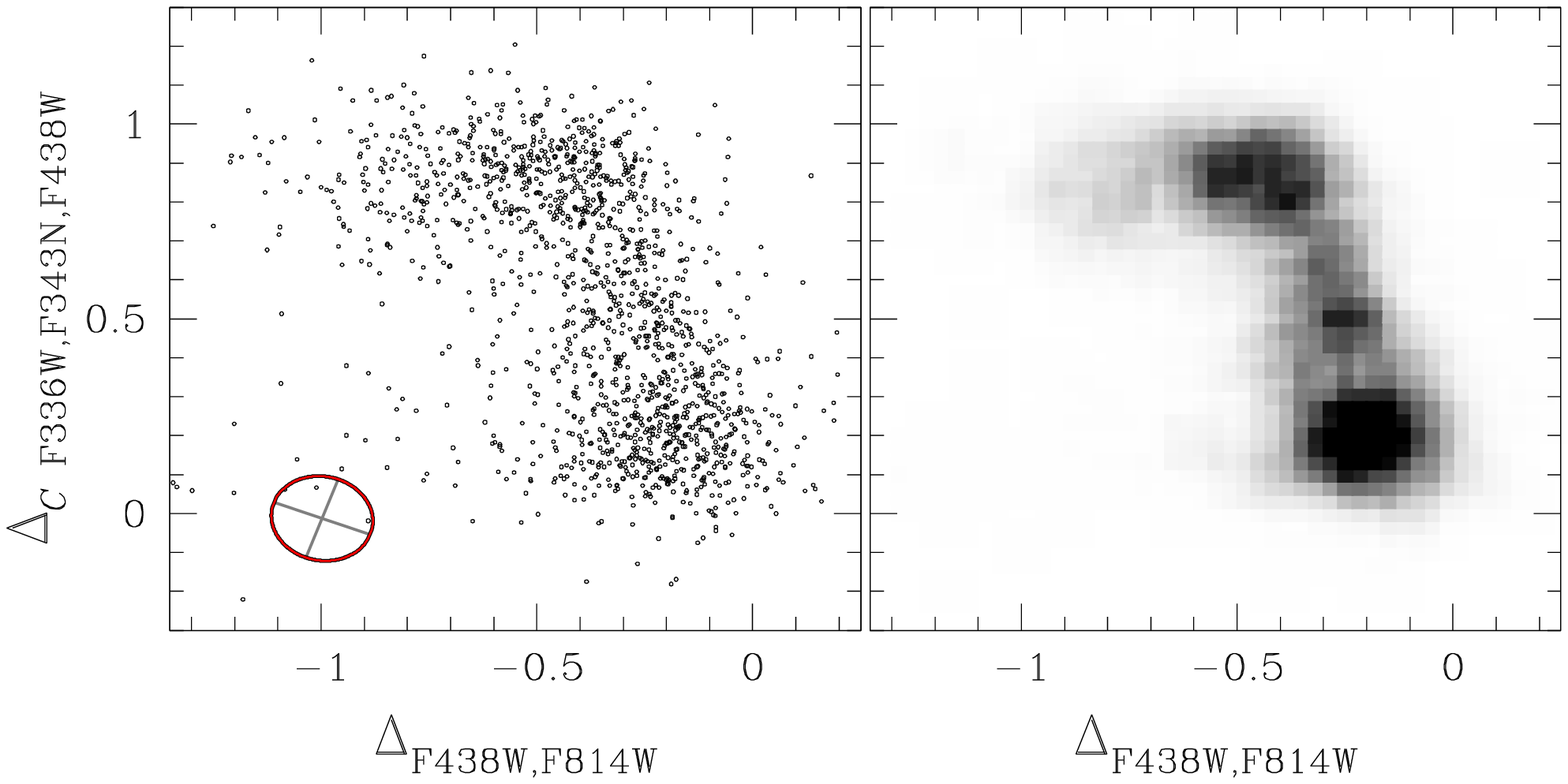}
  \caption{Chromosome map of RGB stars (left panel) and corresponding Hess diagram (right panel). The red ellipse plotted on the left is indicative of the distribution of the observational uncertainties and is derived from artificial stars. It encloses the $68.27\%$ of the simulated ASs.} 
  \label{fig:ChM} 
 \end{figure*} 
 \end{centering}

  The resulting $\Delta_{\rm C F336W,F343N,F438W}$ vs.\,$\Delta_{\rm F438W, F814W}$ ChM is illustrated in the left panel of Fig.~\ref{fig:ChM}, while the corresponding Hess diagram is shown on the right side.
  This figure immediately reveals that NGC\,2419 hosts four main stellar populations clustered around ($\Delta_{\rm F438W,F814W},\Delta_{C \rm F336W,F343N,F438W}$)=($-$0.2,0.2), ($-$0.3,0.5), ($-$0.4,0.9) and ($-$0.8,0.9).  In the next subsections we properly identify these four populations and determine their relative stellar fractions. 
  
\subsection{Distinguishing the four stellar populations}\label{subsec:4p}  
\begin{centering} 
 \begin{figure*} 
  \includegraphics[width=15cm]{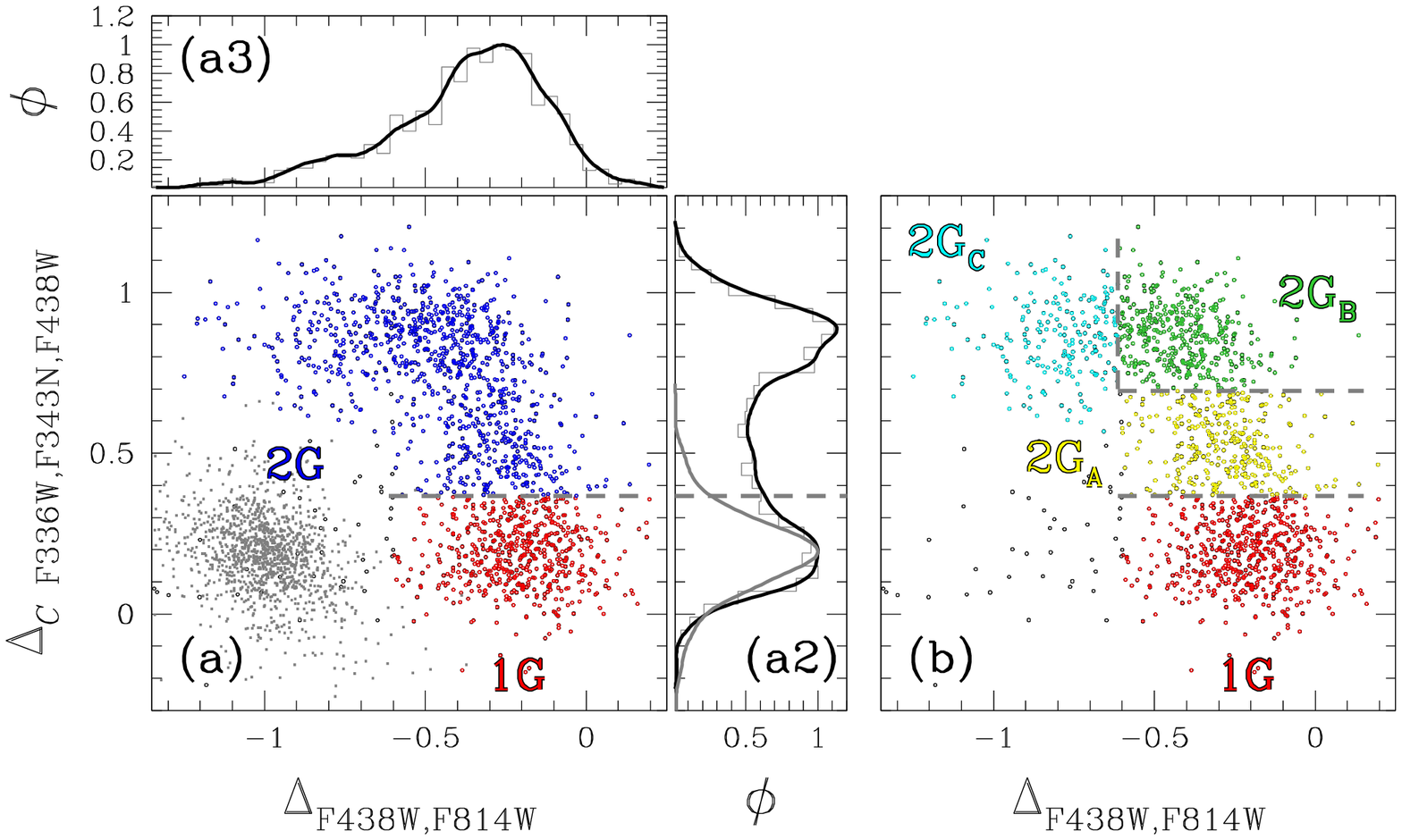}
  \caption{This figure illustrates the procedure that we used to identify the main stellar populations of NGC\,2419. Panel a reproduces the ChM of Fig.~\ref{fig:ChM}. The grey points plotted on the bottom-left corner represent the observational errors. Panels a2 and a3 show the $\Delta_{C \rm F336W,F343N,F438W}$ and $\Delta_{\rm F438W,F814W}$ histogram distributions, respectively, and the corresponding kernel-density distributions (black lines). The gray continuous line plotted in panel a2 is the $\Delta_{C \rm F336W,F343N,F438W}$ kernel-density distribution of the observational uncertainties, while the horizontal dashed line is used to separate bona-fides 1G stars from 2G stars, which are colored red and blue in panel a.
     The four groups of 1G, 2G$_{\rm A}$, 2G$_{\rm B}$, and 2G$_{\rm C}$ are represented with red, yellow, green, and cyan colors, respectively, in panel b. } 
  \label{fig:pops} 
 \end{figure*} 
 \end{centering}

 To identify the main stellar populations of NGC\,2419 we used the procedure illustrated in Fig.~\ref{fig:pops} that is similar to that used by Milone et al.\,(2017, 2018) to define 1G and 2G stars in 58 GCs.

 Panel a of Fig.~\ref{fig:pops} reproduces the ChM of RGB stars plotted in Fig.~\ref{fig:ChM}. We identify 1G stars as those clustered around the origin of the reference frame, while 2G stars are those in the sequence that reaches large values of $\Delta_{C \rm F336W,F343N,F438W}$. Clearly, 2G stars include three stellar populations that we name 2G$_{\rm A}$, 2G$_{\rm B}$, and 2G$_{\rm C}$, with the latter corresponding to the population with the most-extreme $\Delta_{\rm F438W,F814W}$ values. The normalized $\Delta_{C \rm F336W,F343N,F438W}$ and $\Delta_{\rm F438W,F814W}$ histogram distributions are shown in panels a2 and a3.
 
The expected distribution of the photometric errors is represented with grey points, while the corresponding $\Delta_{C \rm F336W,F343N,F438W}$ kernel-density distribution is plotted with a gray line.
The adopted average $\Delta_{\rm F438W,F814W}$ value of the gray points is chosen arbitrarily while the adopted average $\Delta_{C \rm F336W,F343N,F438W}$ value, $\Delta_{C \rm F336W,F343N,F438W}^{0}$, is determined by using the procedure by Milone et al.\,(2018, see their Sect.~2.1).

In a nutshell, we assumed various possible values for $\Delta_{C \rm F336W,F343N,F438W}^{0}$, $\Delta_{C \rm F336W,F343N,F438W}^{i, 0}$, that range from $-$0.200 to 0.100 in steps of 0.001. For each choice of $\Delta_{C \rm F336W,F343N,F438W}^{i, 0}$ we derived the corresponding $\Delta_{C \rm F336W,F343N,F438W}$ kernel-density distribution of the errors, $\phi^{\rm i}_{\rm err}$, and the observed kernel-density distribution, $\phi^{\rm i}_{\rm obs}$.
We compared the distributions $\phi^{\rm i}_{\rm err}$ and $\phi^{\rm i}_{\rm obs}$ for $\Delta_{C \rm F336W,F343N,F438W} < (\Delta_{C \rm F336W,F343N,F438W}^{\rm i, 0} + \sigma)$, where $\sigma$, is defined as the 68.27$^{\rm th}$ percentile of the $\Delta_{C \rm F336W,F343N,F438W}$ distribution of the errors, and calculated the corresponding $\chi$ square. Both distributions are normalized in such a way that their maximum values, calculated in the interval with $\Delta_{C \rm F336W,F343N,F438W} < (\Delta_{C \rm F336W,F343N,F438W}^{\rm i, 0} + \sigma)$, correspond to one.
 We adopted as $\Delta_{C \rm F336W,F343N,F438W}^{0}$ the value of $\Delta_{C \rm F336W,F343N,F438W}^{\rm i, 0}$ that provides the minimum $\chi$ square. 

 The gray dashed horizontal line is plotted at the $\Delta_{C \rm F336W,F343N,F438W}$ level corresponding to the 1.5$\sigma$ deviation from $\Delta_{C \rm F336W,F343N,F438W}^{0}$ and is used to separate the bulk of 1G stars (red points) from 2G stars (blue points). The observed kernel-density distribution of $\Delta_{C \rm F336W,F343N,F438W}$ and the error distribution corresponding to the minimum $\chi$ square are represented with black and gray lines, respectively, in panel a2  of Fig.~\ref{fig:pops}. For completeness, we show the observed $\Delta_{\rm F438W,F814W}$ kernel-density distribution in panel a3.

To identify a sample of bona-fides 2G$_{\rm C}$ stars we extended to 2G stars the procedure described above. In this case we used the distribution of $\Delta_{\rm F438W,F814W}$ to separate the bulk of 2G$_{\rm C}$ stars from the remaining 2G stars.
Finally, we exploit the $\Delta_{C \rm F336W,F343N,F438W}$ to separate the majority of 2G$_{\rm B}$ stars from 2G$_{\rm A}$ stars.

The four groups of 1G, 2G$_{\rm A}$, 2G$_{\rm B}$, and 2G$_{\rm C}$ stars are colored red, yellow, green and cyan, respectively in Fig.~\ref{fig:pops}b and will be used in the next subsection to estimate the relative fraction of stars in each population.

\subsection{Population ratios}
\begin{centering} 
 \begin{figure*} 
  \includegraphics[width=15cm]{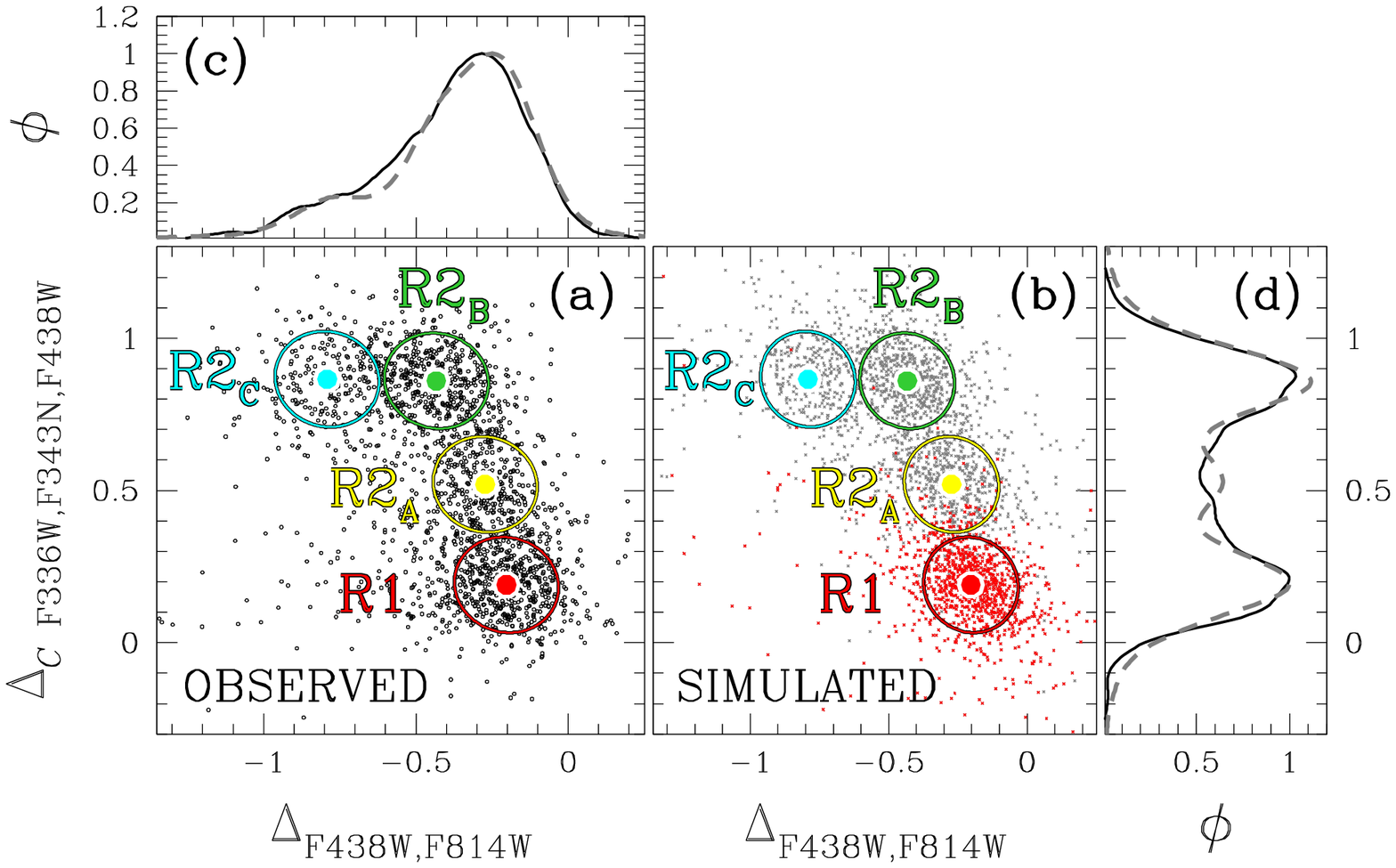}
  \caption{Illustration of the method to estimate the population ratios. Panel a is a reproduction of the observed ChM plotted in Fig.~\ref{fig:ChM} while Panel b shows the simulated ChM. The colored dots are the centers of the four main populations of NGC\,2419 while the four regions, R1, R2$_{\rm A}$, R2$_{\rm B}$, and R2$_{\rm C}$ used to derive the fraction of stars in each population are represented with red, yellow, green, and cyan ellipses, respectively. The ASs used to simulate 1G stars are colored in red. Panels c and d compare respectively the  $\Delta_{\rm F438W,F814W}$ and $\Delta_{C \rm F336W,F343N,F438W}$ kernel-density distributions of observed (black continuous lines) and simulated stars (gray dashed lines).} 
  \label{fig:pratio} 
 \end{figure*} 
 \end{centering}

To estimate the fraction of stars in each population identified in Sect.~\ref{subsec:4p} we extended the method by Milone et al.\,(2012) and Nardiello et al.\,(2018) to the ChM of NGC\,2419.
The procedure is illustrated in Fig.~\ref{fig:pratio}.
Briefly, we calculated the average values of $\Delta_{\rm F438W,F814W}$ and $\Delta_{C \rm F336W,F343N,F438W}$ for the stars of each population (colored dots in Fig.~\ref{fig:pratio}) and used these points as centers of four regions, namely R1, R2$_{\rm A}$, R2$_{\rm B}$, and R2$_{\rm C}$. Each region is an ellipse and is similar to the ellipse that reproduces the distribution of photometric uncertainties.

Due to observational errors, each region includes stars of all the populations. Specifically, the observed number of stars within the region R1 is:
\begin{equation}
N_{\rm R1}=N_{\rm 1G} f_{\rm R1}^{\rm 1G}+ N_{\rm 2G_{\rm A}} f_{\rm R1}^{\rm 2G_{\rm A}}+ N_{\rm 2GB} f_{\rm R1}^{\rm 2G_{\rm B}}+ N_{\rm 2G_{\rm C}} f_{\rm R1}^{\rm 2G_{\rm C}}
\end{equation}
 where $N_{\rm 1G~(2G_{\rm A}, 2G_{\rm B}, 2G_{\rm C})}$ is the total number of analyzed 1G (2G$_{\rm A}$, 2G$_{\rm B}$, 2G$_{\rm C}$) stars and $f_{\rm R1}^{\rm 1G~(2G_{\rm A}, 2G_{\rm B}, 2G_{\rm C})}$ is the fraction of 1G (2G$_{\rm A}$, 2G$_{\rm B}$, 2G$_{\rm C}$) stars in the region R1. 
 The number of observed stars in the regions R2$_{\rm A}$, R2$_{\rm B}$, and R2$_{\rm C}$ are related to the fractions of stars of each population by three similar equations.

 By solving these four equations we calculate the total numbers of 1G, 2G$_{\rm A}$, 2G$_{\rm B}$, and 2G$_{\rm C}$ stars and find that the corresponding fraction of stars with respect to the total number of analyzed RGB stars are 37$\pm$1\%, 20$\pm$1\%, 31$\pm$1\% and 12$\pm$1\%, respectively. 
 As a consequence, the whole 2G comprises 63$\pm$1\% of the total number of analyzed stars.

\begin{centering} 
 \begin{figure*} 
  \includegraphics[width=12.00cm]{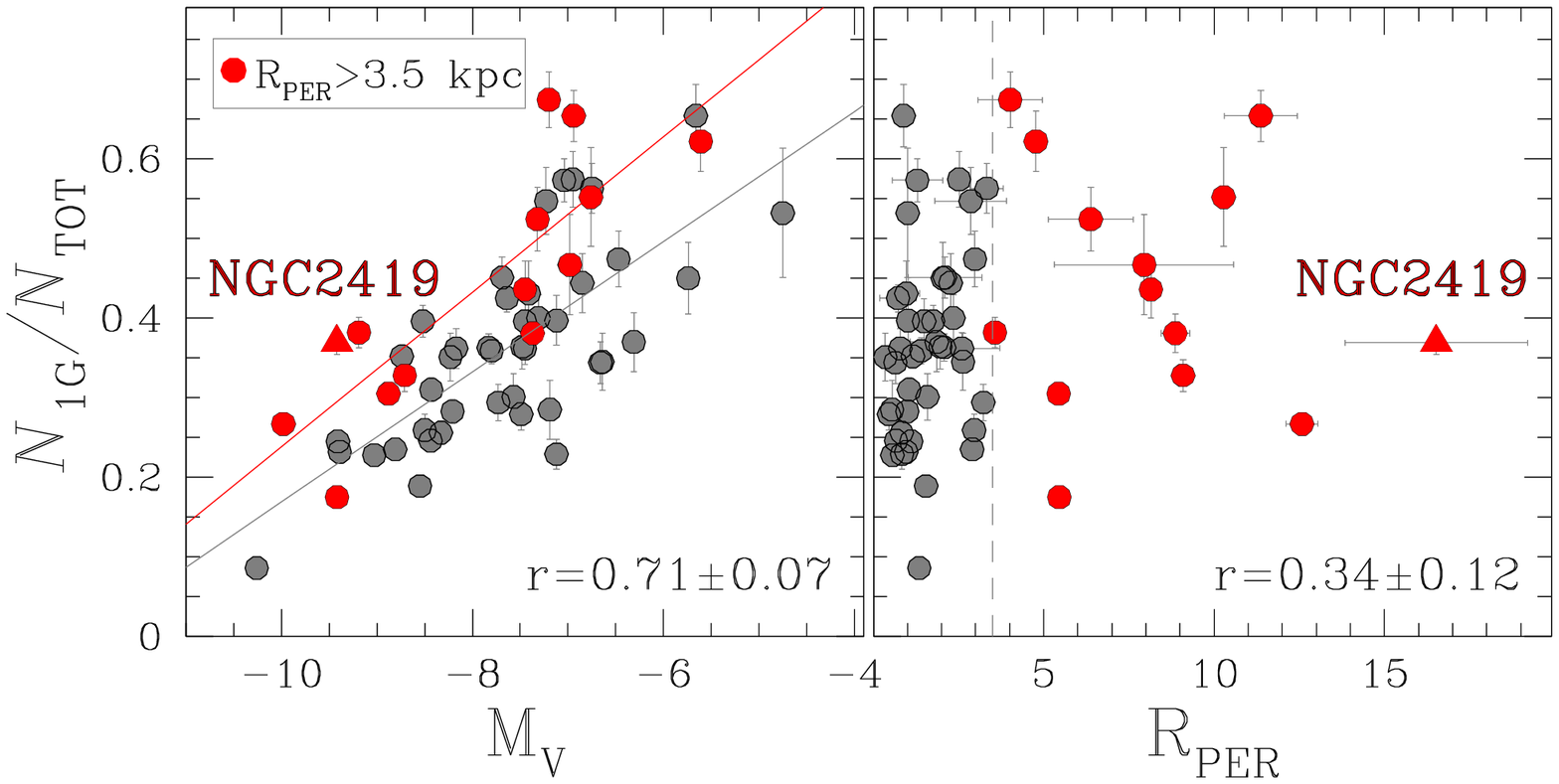}
  \caption{Fraction of 1G stars with respect to the total number of analyzed stars as a function of the absolute magnitude (from Harris 1996, left panel) and the perigalactic distance (from Baumgardt et al.\,2019, right panel). Circles indicate the clusters analyzed by Milone et al.\,(2017, 2018) while NGC\,2419 is marked with triangles. Clusters with perigalactic distance larger than 3.5 kpc (vertical dashed line in the right panel) are colored in red, while gray dots represent the remaining clusters. The corresponding least-squares best-fit straight lines are shown in the left panel and indicate that GCs with R$_{\rm PER} >3.5$ kpc have on average larger fraction of 1G stars that the remaining GCs at a given luminosity.} 
  \label{fig:G1vsP} 
 \end{figure*} 
 \end{centering}

 To compare NGC\,2419 with the other GCs we plot in the left panel of Fig.~\ref{fig:G1vsP} the fraction of 1G stars as a function of the absolute visual magnitude, $M_{\rm V}$ (from Harris 1996, updated as in 2010), for the 58 clusters studied by Milone et al.\,(2017, 2018) and for NGC\,2419.
  The fraction of 1G stars correlates with $M_{\rm V}$ with a Spearman's rank correlation coefficient, r=0.71$\pm$0.07.
  However, although NGC\,2419 follows the general trend, its fraction of 1G stars is larger than that of most clusters with similar luminosity.

  The right panel of Fig.~\ref{fig:G1vsP} shows that the fraction of 1G stars does not exhibit significant correlation with the perigalactic distances, R$_{\rm PER}$, of the host GC from (from Baumgardt et al.\,2019, r=0.34$\pm$0.12). Similarly, we verify that there is no significant correlation with the apogalactic distance (from Baumgardt et al.\,2019, r=0.31$\pm$0.13) and with the distance from the Galactic center (from the 2010 version of the Harris\,1996 catalog, r=0.00$\pm$0.13).
   However, we note that the clusters with large perigalactic distances R$_{\rm PER}>3.5$ kpc, exhibit, on average, larger fractions of 1G than those of the remaining clusters with similar masses as shown in the left panel of Fig.~\ref{fig:G1vsP}.

 \section{The chemical composition of the stellar populations}\label{sec:He}
 To infer the chemical composition of the four stellar populations of NGC\,2419 we combine photometry from this paper and chemical abundances inferred from spectroscopy in the literature.
 Specifically, we exploit the results by Mucciarelli et al.\,(2012) who analyzed 49 giants by using medium-resolution spectra collected with the multi-object spectrograph DEIMOS@Keck.
 They find that NGC\,2419 has homogeneous abundances of Fe, Ca, and Ti and discovered large star-to-star variations in Mg and K.

\begin{centering} 
 \begin{figure*} 
  \includegraphics[height=6cm]{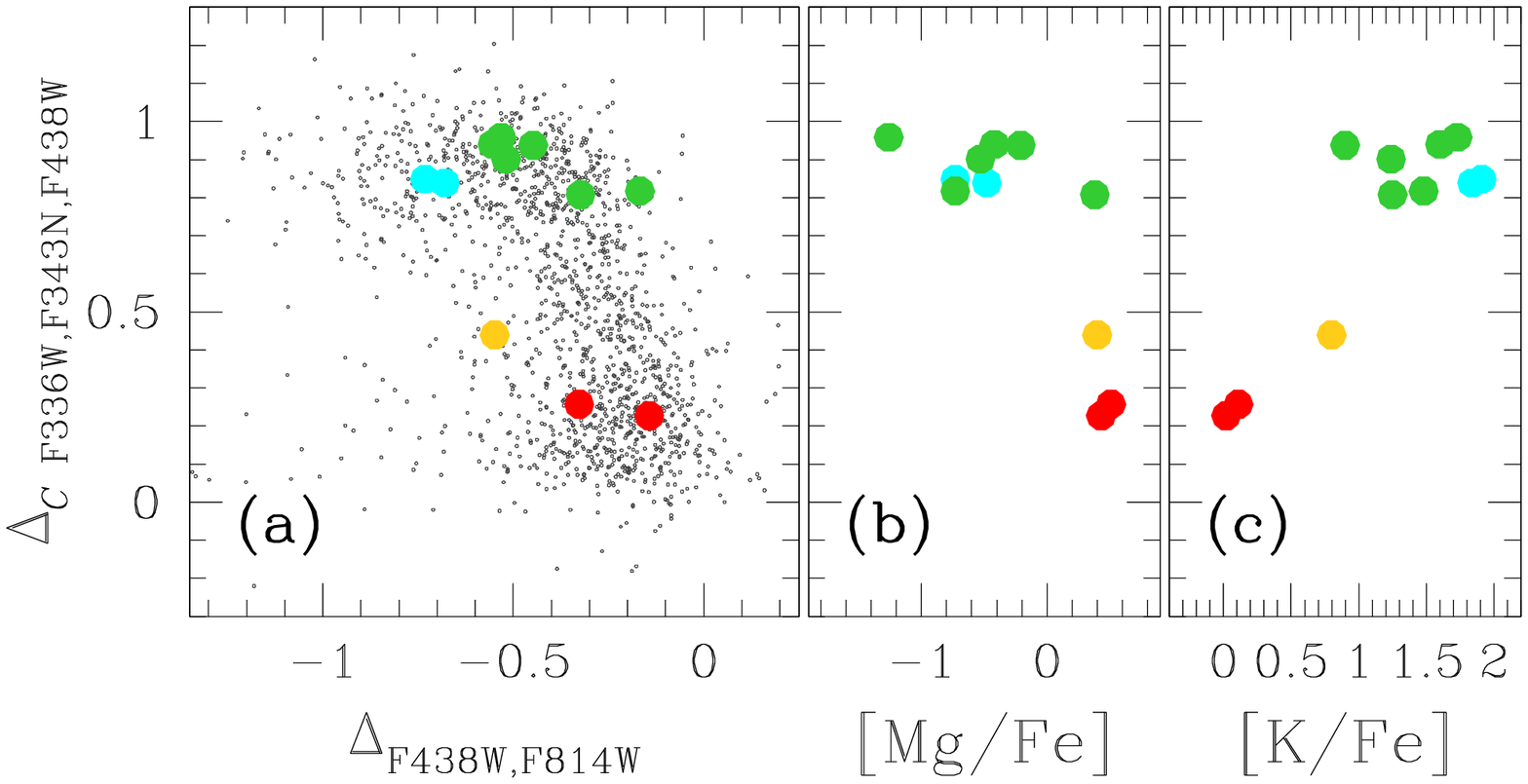}
  \includegraphics[height=6cm]{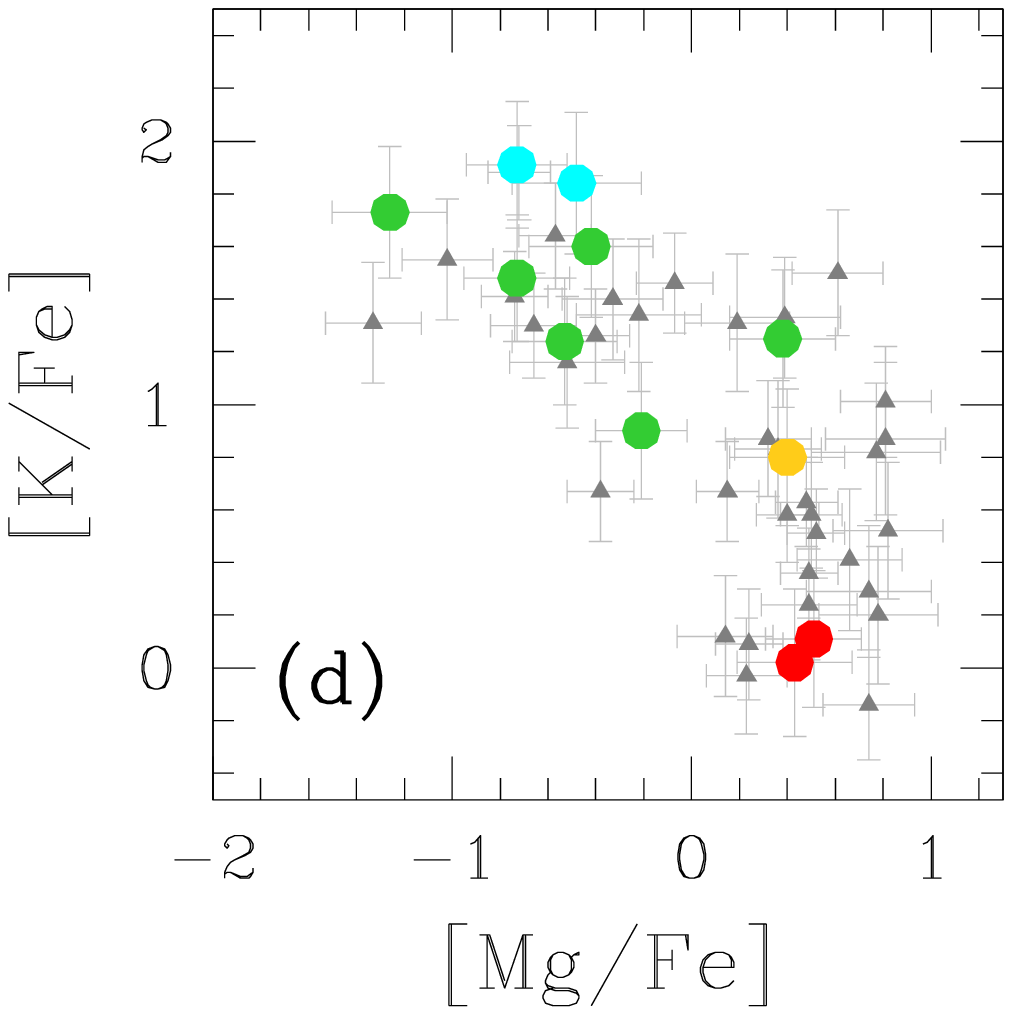}
  \caption{$\Delta_{C \rm F336W,F343N,F438W}$ vs.\,$\Delta_{\rm F438W,F814W}$ ChM of RGB stars in NGC\,2419 (panel a). The stars in the ChM for which spectroscopy by Mucciarelli et al.\,(2012) is available are marked with large dots in all the panels. 1G, 2G$_{\rm A}$, 2G$_{\rm B}$ and 2G$_{\rm C}$ stars are colored with red, yellow, green and cyan, respectively. Panel b and c show $\Delta_{C \rm F336W,F343N,F438W}$ against [Mg/Fe] and [K/Fe], respectively, while panel d reproduces the potassium-magnesium anticorrelation from Mucciarelli et al.\,(2012).} 
  \label{fig:abb} 
 \end{figure*} 
 \end{centering}

 Chemical abundances from Mucciarelli and collaborators are available for eleven stars in the ChM of Fig.~\ref{fig:ChM}, including two 1G stars, one 2G$_{\rm A}$ star, six  2G$_{\rm B}$ stars and two 2G$_{\rm C}$ stars (Fig.~\ref{fig:abb}a). 
 Panels b and c of Fig.~\ref{fig:abb} show that $\Delta_{C \rm F336W,F343N,F438W}$ anticorrelates with [Mg/Fe] and correlates with [K/Fe] while panel d of Fig.~\ref{fig:abb} reproduces the potassium-magnesium anticorrelation from Mucciarelli et al.\,(2012).
 
 The comparison between the ChM and the chemical abundances indicates that 1G and 2G$_{\rm A}$ stars have similar magnesium abundance of [Mg/Fe]$\sim$0.4, while 2G$_{\rm B}$ and 2G$_{\rm C}$ are depleted in magnesium by $\sim$1 dex with respect to the remaining stars of NGC\,2419. 1G stars have nearly solar potassium abundance, while the 2G$_{\rm A}$ is enhanced in [K/Fe] by $\sim$0.8 dex. The remaining stars exhibit extreme potassium contents up to [K/Fe]$\sim$1.4 dex for the six 2G$_{\rm B}$ stars and [K/Fe]$\gtrsim$1.9 for the two 2G$_{\rm C}$ stars. As discussed by Mucciarelli et al.\,(2012) there is no evidence for significant variations of Ca, Fe, and Ti. The average elemental abundances for each population are provided in Table~\ref{tab:abb}.
 
\begin{table*}
  \caption{Average abundance and corresponding random mean scatter (rms) of Mg, K, Ca, Fe, and Ti for 1G, 2G, 2G$_{\rm B}$, and 2G$_{\rm C}$ stars.
    We also list the abundances  and the corresponding uncertainties of the only 2G$_{\rm A}$ for which chemical abundances are available. The number, $N$, of stars used to derive the quoted abundances of each population are also indicated.
     The elemental abundances are taken from Mucciarelli et al.\,(2012).}
\begin{tabular}{l ccc ccc ccc ccc ccc}
\hline \hline
       &        &   1G &    &          &  2G    &     &         & 2G$_{\rm A}$&       &     &2G$_{\rm B}$&        &     &2G$_{\rm C}$& \\           
       & Average &rms & $N$ & Abundance&$\sigma$& $N$ & Average & rms & $N$ &Average&rms& $N$ &Average&rms& $N$ \\
\hline   
    [Mg/Fe]&   0.47 & 0.06 & 2  &$-$0.40 & 0.53 & 9 &     0.40 & 0.24 & 1 &   $-$0.46 & 0.55 & 6 &   $-$0.61 & 0.18 & 2 \\
    
    [K/Fe] &   0.07 & 0.06 & 2  &   1.42 & 0.40 & 9 &     0.80 & 0.26 & 1 &      1.37 & 0.30 & 6 &      1.88 & 0.05 & 2 \\

    [Ca/Fe]&   0.31 & 0.24 & 2  &   0.50 & 0.06 & 9 &     0.51 & 0.07 & 1 &      0.49 & 0.08 & 6 &      0.53 & 0.00 & 2 \\

    [Fe/H] &$-$1.98 & 0.21 & 2  &$-$2.08 & 0.10 & 9 &  $-$2.24 & 0.13 & 1 &   $-$2.06 & 0.08 & 6 &   $-$2.05 & 0.97 & 2 \\ 

    [Ti/Fe]&   0.14 & 0.06 & 2  &   0.29 & 0.10 & 9 &     0.34 & 0.21 & 1 &      0.28 & 0.12 & 6 &      0.28 & 0.01 & 2 \\
     \hline\hline
\end{tabular}
  \label{tab:abb}
 \end{table*}

\subsection{The Helium abundance of multiple stellar populations}

 \begin{centering} 
 \begin{figure*} 
  \includegraphics[width=15cm]{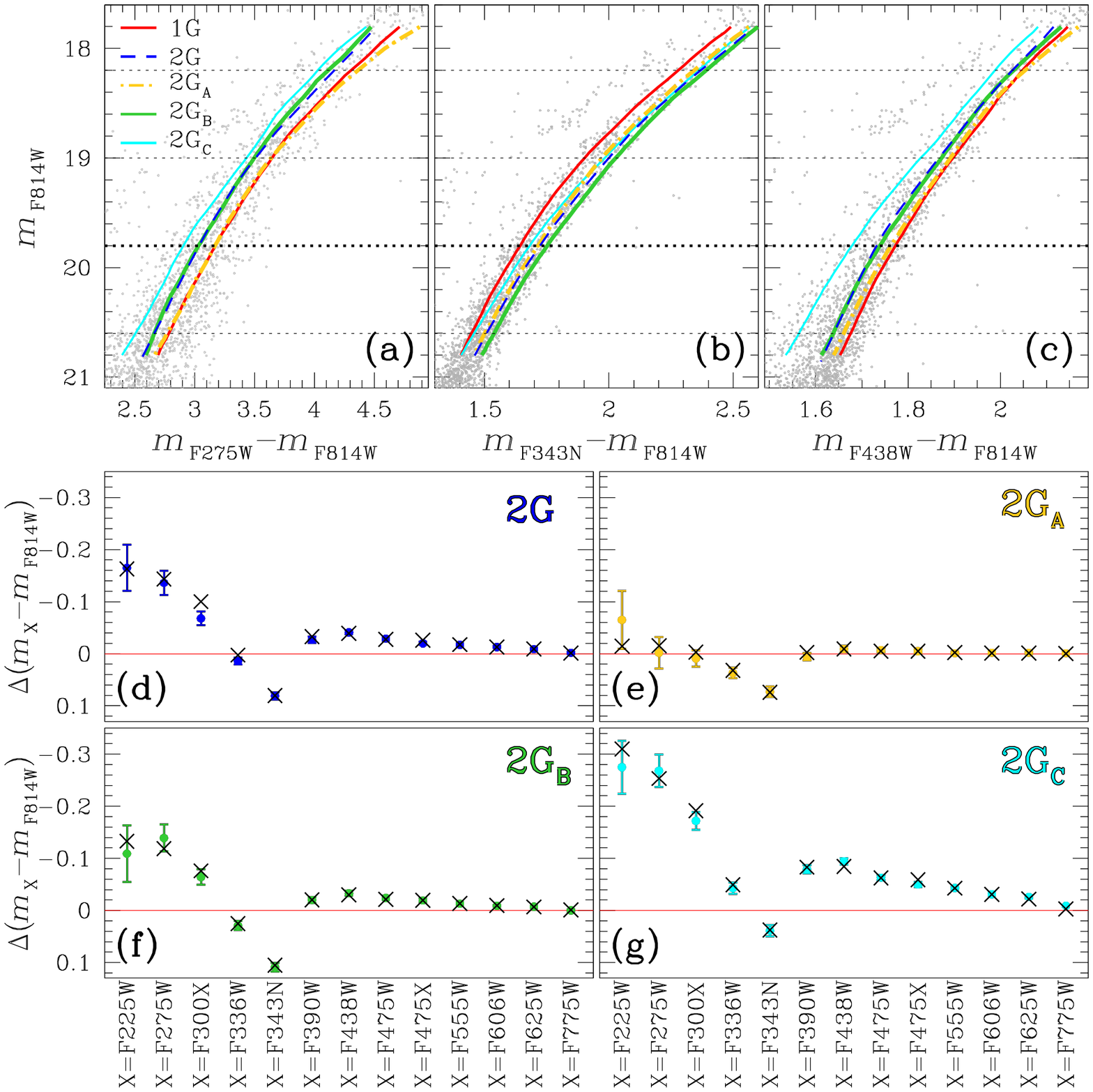}
  \caption{\textit{Upper panels.} Zoom of the $m_{\rm F814W}$ vs.\,$m_{\rm F275W}-m_{\rm F814W}$ (panel a), $m_{\rm F814W}$ vs.\,$m_{\rm F343N}-m_{\rm F814W}$ (panel b) and $m_{\rm F814W}$ vs.\,$m_{\rm F438W}-m_{\rm F814W}$ (panel c) CMDs around the RGB. The colored lines overimposed on each CMD are the fiducials of the stellar populations identified in the paper, while the dotted horizontal lines correspond to four values of $m_{\rm F814W}^{\rm cut}$ used to infer the helium abundance of each population. \textit{Lower panels.} The colored dots are the observed $m_{\rm X}-m_{\rm F814W}$ color difference between the fiducial of 2G (panel d), 2G$_{\rm A}$ (panel e), 2G$_{\rm B}$ (panel f), and 2G$_{\rm C}$ (panel g), and the fiducial of 1G stars for various X filters calculated for $m_{\rm F814W}^{\rm ref}=19.8$. The colors inferred from the best-fit synthetic spectrum are represented with black crosses.} 
  \label{fig:He} 
 \end{figure*} 
 \end{centering} 

We estimated the relative helium abundance between 2G and 1G stars ($\delta$Y$_{\rm 2G,1G}$) 
 by following the method by Milone et al.\,(2012, 2018) that is illustrated in Fig.~\ref{fig:He}. 
 In a nutshell,  we first derived the RGB fiducial lines of 1G and 2G stars in the CMDs $m_{\rm F814W}$ vs.\,$m_{\rm X}-m_{\rm F814W}$, where X=F225W, F275W, F300X, F336W, F343N, F390W, F438W, F475W, F475X, F555W, F606W, F625W and F775W. As an example, in the panels a, b, and c of Fig.~\ref{fig:He} we show the fiducial lines corresponding to X=F275W, X=F343N, and X=F438W, respectively.
We find that the 2G fiducials have bluer colors than the 1G fiducials in all the CMDs but for X=F336W and F343N, where 2G stars are redder than 1G stars.  
 
 We defined four equally-spaced reference points in the magnitude bin with $18.0<m_{\rm F814W}<20.8$ and calculated the $m_{\rm X}-m_{\rm F814W}$ color difference  between the fiducial of 2G and 1G stars for each reference point, $\Delta$($m_{\rm X}-m_{\rm F814W}$). The magnitude levels corresponding to these reference points ($m_{\rm F814W}^{\rm ref}=18.2, 19.0, 19.8, 20.6$) are represented with dotted lines in panels a, b, and c of Fig.~\ref{fig:He} while in panel d we represent with colored dots the $\Delta$($m_{\rm X}-m_{\rm F814W}$) values corresponding to the various X filters for the reference point with $m_{\rm F814W}^{\rm ref}=19.8$.  

 We derived the gravity and effective temperature of 1G stars with luminosities that correspond to the reference points by using the best-fit isochrones from the Darthmouth database (Dotter et al.\,2008). We assumed primordial helium content, Y=0.246, iron abundance, [Fe/H]=$-$2.09 (Mucciarelli et al.\,2012), and [$\alpha$/Fe]=0.40.  The best fit between the isochrones and the data is provided by distance modulus and reddening of (m$-$M)$_{0}=$19.68 and E(B$-$V)=0.07, respectively, and age of 13.0 Gyr, which are similar to the values listed by Harris (1996, 2010 update) and Dotter et al.\,(2010).

 For each reference point we calculated a synthetic spectrum and a grid of comparison spectra with different chemical composition by using the computer programs ATLAS 12 and Synthe (Kurucz 2005; Castelli et al.\,2005; Sbordone et al.\,2007).  
  We assumed [C/Fe]=$-$0.6, [N/Fe]=0.6, [O/Fe]=0.4, as inferred from high-resolution spectra of 1G stars in the metal-poor GC M\,22 by Marino et al.\,(2011), and the values of effective temperature and gravity that we derived from the best-fit isochrone and are provided in Tab.~\ref{tab:He}. The comparison spectra have [C/Fe] that ranges between $-$1.5 and 0.0 dex and [O/Fe] that ranges from $-1.0$ to 0.4 in steps of 0.1 dex. [N/Fe] varies from 0.30 to 2.00 in steps of 0.05 dex. The helium content of the comparison spectra ranges from Y=0.246 to 0.470 in steps of 0.001 and the values of effective temperature and gravity are derived from the corresponding isochrone from Dotter et al.\,(2008). The magnesium content is fixed and corresponds to the average [Mg/Fe] abundances of 1G and 2G stars inferred from the data by Mucciarelli et al.\,(2012) and listed in Table~\ref{tab:abb}.

 The corresponding color differences have been derived from the convolution of each spectrum with the transmission curves of the WFC3/UVIS filters used in this paper. The best determinations of the relative helium content between 2G and 1G star are given by the chemical composition of the comparison spectrum that provides the best match with the observed color differences and correspond to an helium difference $\delta$Y$_{\rm 2G-1G}$=0.07$\pm$0.01, where the uncertainty is estimated as the rms of the four helium determinations divided by the square root of three. As an example, the black crosses plotted in panel d of Fig.~\ref{fig:He} represent the color differences corresponding to the spectra that provide the best fit with the observed color difference between 2G and 1G stars for $m_{\rm F814W}^{\rm ref}=19.8$. 
 
 The procedure described above has been also used to infer the relative abundances of He between 2G$_{\rm A}$, 2G$_{\rm B}$, 2G$_{\rm C}$ and 1G stars.
 Results are listed in Table~\ref{tab:He} for each value of $m_{\rm F814W}^{\rm ref}$, while in panels f, g, and e of  Fig.~\ref{fig:He} we show the observed difference between the color of each population and the color of 1G stars for $m_{\rm F814W}^{\rm ref}=19.8$ and the corresponding color differences inferred from the best-fit spectra.

\begin{centering} 
 \begin{figure} 
  \includegraphics[width=8.5cm]{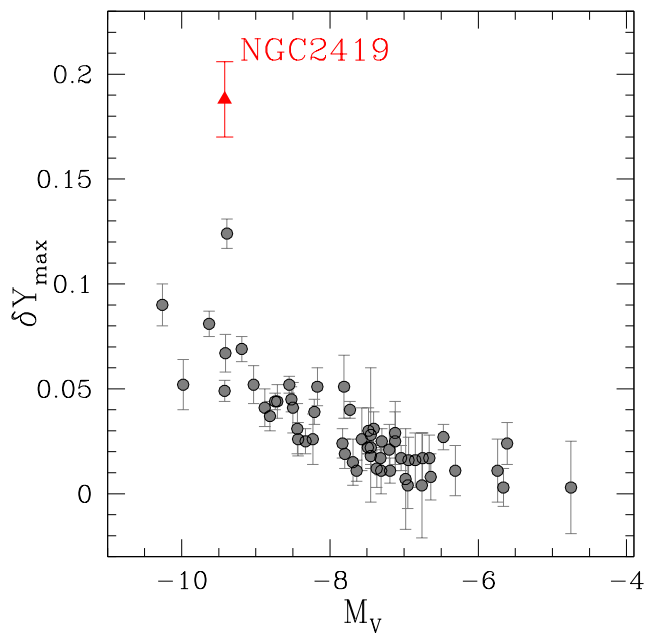}
  \caption{Maximum helium variation as a function of the absolute magnitude of the host GC.} 
  \label{fig:HevsMv} 
 \end{figure} 
 \end{centering}

 We conclude that 2G$_{\rm A}$, 2G$_{\rm B}$, 2G$_{\rm C}$ are enhanced in helium mass fraction by 0.01$\pm$0.01, 0.06$\pm$0.01, and 0.19$\pm$0.02 dex, respectively, with respect to 1G stars. Figure~\ref{fig:HevsMv} compares the maximum helium variation that we derived for NGC\,2419 with results for 58 GCs by Milone et al.\,(2018).
 
The procedure that we adopted to infer the helium content of the stellar populations also allows to constrain the relative abundances of C, N, and O (see Milone et al.\,2015, 2017 for details). From the best-fit spectra we find that 2G$_{\rm A}$ are enhanced in nitrogen by 0.3$\pm$0.1 dex and depleted in C and O by 0.3$\pm$0.3 and 0.2$\pm$0.2 dex, respectively, with respect to 1G stars. 2G$_{\rm B}$ stars exhibit higher [N/Fe] than the 1G (by 0.7$\pm$0.1 dex), and  lower [C/Fe] and [O/Fe] (by 0.6$\pm$0.2 and 0.5$\pm$0.1 dex, respectively).  The 2G$_{\rm C}$ is enhanced in nitrogen by 0.7$\pm$0.1 dex and depleted in carbon and oxygen by 0.7$\pm$0.3 and 0.6$\pm$0.2 dex, respectively, with respect to 1G stars.

\begin{table*}
  \caption{Atmospheric parameters and helium variations of the best-fit spectra inferred for the four reference magnitudes and average helium variations.}
\begin{tabular}{ccc ccc ccc ccc ccc}
\hline \hline
                  &  1G  &     &      &       2G &  &      &       2G$_{\rm A}$ & &      &        2G$_{\rm B}$  &    &      & 2G$_{\rm C}$   &      \\
\hline
$m_{\rm F814W}^{\rm ref}$&   $T_{\rm eff}$   &  $\log{g}$    &  $T_{\rm eff}$   &  $\log{g}$  &  $\delta$Y&  $T_{\rm eff}$   &  $\log{g}$  &  $\delta$Y&  $T_{\rm eff}$   &  $\log{g}$  &  $\delta$Y&  $T_{\rm eff}$   &  $\log{g}$  &  $\delta$Y\\
\hline
18.2  & 4844 & 1.73 & 4881 & 1.69 & 0.070 & 4842 & 1.73 & $-$0.003 & 4871 & 1.70 & 0.052 & 4926 & 1.62 & 0.173 \\  
19.0  & 5006 & 2.10 & 5038 & 2.07 & 0.055 & 5006 & 2.10 &    0.000 & 5033 & 2.07 & 0.047 & 5086 & 2.00 & 0.153 \\
19.8  & 5157 & 2.47 & 5206 & 2.42 & 0.078 & 5162 & 2.47 &    0.008 & 5139 & 2.44 & 0.058 & 5274 & 2.33 & 0.208 \\ 
20.6  & 5286 & 2.82 & 5340 & 2.79 & 0.073 & 5298 & 2.81 &    0.017 & 5331 & 2.79 & 0.061 & 5436 & 2.68 & 0.219 \\
\hline
          &                      &     &     &                        &       & &     Average             &       & &                        &     &  &    &  \\
\hline
          & 2G &  &     & 2G$_{\rm A}$ &   & &   2G$_{\rm B}$ &  &  & 2G$_{\rm C}$ &  &  &    &    \\
          & $\delta$Y & rms &     & $\delta$Y & rms   & &   $\delta$Y & rms  &  & $\delta$Y & rms  &  &    &    \\
\hline                                                                                                                                                                     
          &  0.069                & 0.010 &  & 0.006                  & 0.009 & &   0.055                 & 0.006 &  & 0.188 & 0.031                &  &    &    \\    
     \hline\hline
\end{tabular}
  \label{tab:He}
 \end{table*}

  
  

\section{Summary and discussion}\label{sec:discussion}
We exploited {\it HST} images  of the massive outer-halo GC NGC\,2419 to investigate its multiple stellar populations by using photometry in fourteen bands of UVIS/WFC3.  The $\Delta_{ C \rm F336W,F343N,F438W}$ vs.\,$\Delta_{\rm F438W,F814W}$ ChM of RGB stars 
 reveals that NGC\,2419 hosts four main stellar populations of 1G, 2G$_{\rm A}$, 2G$_{\rm B}$ and 2G$_{\rm C}$ stars that comprise the 37$\pm$1\%, 20$\pm$1\%, 31$\pm$1\% and 12$\pm$1\% of stars, respectively, of the total number of analyzed stars. 

 Milone et al.\,(2017, 2018) estimated the relative numbers of 1G and 2G stars in 58 Galactic GCs and find a significant anti-correlation between the fraction 1G stars and the mass of the host GC. The fraction of 1G stars in NGC\,2419, (M$_{\rm V}=-9.42$, Harris 1996 updated as in 2010) is larger than that of all the massive GCs with M$_{\rm V}<-9.0$, the only exception is NGC\,7078 (M$_{\rm V}=-9.19$, Harris 1996 updated as in 2010), where 1G stars comprise 0.40$\pm$0.02\% of the total number of stars.

 Some scenarios on the formation and evolution of multiple populations predict that GCs were dominated by 1G stars at the formation and have lost a large number stars into the Galactic halo corresponding to $\sim$90\% of the total cluster mass. In particular, since 2G stars formed in the innermost cluster regions, the primordial GCs preferentially lost 1G stars (e.g.\,D'Ercole et al.\,2008, 2010; D'Antona et al.\,2016).
 The possibility that NGC\,2419  evolved in isolation and is almost not affected by the tidal influence of the Galaxy suggests that it was not significantly affected by mass-loss due to tidal stripping.

 The presence of a fraction of 1G stars that is larger than that of most GCs with similar masses makes it tempting to speculate that the interaction between the cluster and the Galaxy can affect its present-day ratio between 2G and 1G stars. Nevertheless, the evidence that the 2G includes more than 60\% of the total number of cluster stars of NGC\,2419 is a challenge for the scenarios mentioned above that predict for this isolated cluster a fraction of 1G of $\sim$0.9 (e.g.\,Di Criscienzo et al.\,2011). 

 By comparing the results for NGC\,2419 and for the 58 GCs homogeneously analyzed by Milone et al.\,(2017, 2018) we find no evidence for a strong correlation between the fraction of 1G stars and neither the distance from the Galactic center nor with the perigalactic and the apogalactic distance. Nevertheless, we notice that clusters with large perigalactic distances host, on average, larger fractions of 1G stars than the remaining GCs. This fact suggests that the tidal interactions between the clusters and the Milky Way affects the present-day fraction of 1G stars.
 
Spectroscopic investigation has revealed that NGC\,2419 exhibits extreme star-star-abundance variation in Mg and K (Cohen et al.\,2011, 2012) with an extended Mg-K anticorrelation (Mucciarelli et al.\,2012). Such chemical composition is different from that of the majority of GCs that have homogeneous content of potassium (e.g.\,Carretta et al.\,2013).
From the analysis of eleven stars in the ChM that have been studied by Mucciarelli and collaborators by using medium-resolution HIRES@Keck spectra, we find that the abundance of potassium increases from [K/Fe]$\sim 0.1$ to [K/Fe]$\sim 1.9$ when moving from 1G to 2G$_{\rm C}$ stars. Moreover, 2G$_{\rm B}$ and 2G$_{\rm C}$ stars are significantly depleted in Mg, by $\sim 0.8$ and $0.9$ dex, respectively, with respect to 1G and 2G$_{\rm A}$ stars that have both [Mg/Fe]$\sim$0.5.
 
 Previous evidence of a broad RGB is provided by Di Criscienzo et al.\,(2011b, see their Fig.~9) and Lee et al.\,(2013) and is based on photometry in the F475W and F814W bands collected with WFC/ACS. From the comparison of the observed CMD and isochrones, these authors concluded that NGC\,2419 hosts stars that are heavily enhanced in helium by $\Delta$Y$\sim$0.17-0.19. Di Criscienzo et al.\,(2015), based on the HB of NGC\,2419 suggested a smaller helium enhancement of $\Delta$Y$\sim$0.11. The presence of stellar populations with large helium variations is qualitatively confirmed by Beccari et al.\,(2013) on the basis of the color broadening in the $V$ vs.\,$V-I$ CMD derived from data collected with the Large Binocular Telescope.
 
To further investigate the chemical composition of the four stellar populations of NGC\,2419 and infer their relative helium abundances, we compared the observed colors of RGB stars with the colors derived from synthetic spectra with appropriate chemical composition. 
We find that 2G$_{\rm A}$, 2G$_{\rm B}$ and 2G$_{\rm C}$ stars are respectively enhanced in helium mass fraction by $\delta$Y=0.01$\pm$0.01, 0.06$\pm$0.01, and 0.19$\pm$0.02, with respect to 1G stars that are assumed at primordial helium content (Y$\sim$0.246). 
The extreme helium abundance of 2G$_{\rm C}$ stars that we inferred from multi-band photometry is consistent with the findings by Di Criscienzo et al.\,(2012) and Lee et al.\,(2013).

 Recent work have inferred the helium abundance of large sample of Galactic and extragalactic GCs (Milone et al.\,2018; Lagioia et al.\,2018a,b) and concluded that the internal helium variations is typically smaller than $\sim$0.12 in helium mass fraction. 
Moreover, the maximum internal helium variation correlates with the mass of the host GC (e.g.\,Milone 2015; Milone et al.\,2018).
 The fact that NGC\,2419, which is one of the most-massive GCs of the Milky Way, is the cluster with the largest observed helium variation corroborates the conclusion that the complexity of multiple populations increases with cluster mass. 

 Several scenarios for the formation of multiple populations in GCs proposed that 2G stars formed by the ejecta of more-massive 1G stars. The nature of the polluters is still widely debated. AGB stars with masses of $\sim 3-8 {\mathcal M}_{\odot}$ (e.g.\,Ventura et al.\,2001; D'Antona et al.\,2002, 2016; Tailo et al.\,2015), fast-rotating massive stars (FRMSs, Decressin et al.\,2007; Krause et al.\,2013), and supermassive stars (Denissenkov et al.\,2014) are considered as possible candidates (see Renzini et al.\,2015 for a critical review).
 
 Stars with extreme helium abundance are a challenge for the AGB scenarios because they predict that the maximum helium content of 2G stars is smaller than Y$\sim 0.38$, although Karakas et al.\,(2014) suggested that such stars with extreme helium content can form from the ejecta of a previous generation of helium-rich AGB stars. 
On the contrary, the presence of stars with extreme helium abundance is consistent with the FRMS and super-massive stars scenarios (e.g.\,Prantzos et al.\,2017). As an example, Chantereau et al.\,(2016), based on the FRMS scenario, predict that about 10\% of present-day GC stars have Y$> 0.40$, which is qualitatively consistent with what we observe in NGC\,2419. Nevertheless, the lack of stars with very high helium content (Y$>$0.50) is in contrast with the predictions by Chantereau and collaborators.

\section*{acknowledgments} 
\small 
This work has received funding from the European Research Council (ERC) under the European Union's Horizon 2020 research innovation programme (Grant Agreement ERC-StG 2016, No 716082 'GALFOR', PI: Milone), and the European Union's Horizon 2020 research and innovation programme under the Marie Sklodowska-Curie (Grant Agreement No 797100, beneficiary: Marino). APM and MT acknowledge support from MIUR through the FARE project R164RM93XW ‘SEMPLICE’ (PI: Milone). 
\bibliographystyle{aa}

\begin{thebibliography}{}

\bibitem[Anderson et al.(2008)]{2008AJ....135.2055A} Anderson, J., Sarajedini, A., Bedin, L.~R., et al.\ 2008, \aj, 135, 2055 

\bibitem[Bastian et al.(2013)]{2013MNRAS.436.2398B} Bastian, N., Lamers, H.~J.~G.~L.~M., de Mink, S.~E., et al.\ 2013, \mnras, 436, 2398 
  
\bibitem[Baumgardt et al.(2019)]{2019MNRAS.482.5138B} Baumgardt, H., Hilker, M., Sollima, A., \& Bellini, A.\ 2019, \mnras, 482, 5138 
  
\bibitem[Beccari et al.(2013)]{2013MNRAS.431.1995B} Beccari, G., Bellazzini, M., Lardo, C., et al.\ 2013, \mnras, 431, 1995 
  
\bibitem[Bedin et al.(2005)]{2005MNRAS.357.1038B} Bedin, L.~R., Cassisi, S., Castelli, F., et al.\ 2005, \mnras, 357, 1038 

\bibitem[Bedin et al.(2009)]{2009ApJ...697..965B} Bedin, L.~R., Salaris, M., Piotto, G., et al.\ 2009, \apj, 697, 965
  
\bibitem[Bellini \& Bedin(2009)]{2009PASP..121.1419B} Bellini, A., \& Bedin, L.~R.\ 2009, \pasp, 121, 1419
  
\bibitem[Bellini et al.(2011)]{2011PASP..123..622B} Bellini, A., Anderson, J., \& Bedin, L.~R.\ 2011, \pasp, 123, 622   

\bibitem[Bellini et al.(2017)]{2017ApJ...842....6B} Bellini, A., Anderson, J., Bedin, L.~R., et al.\ 2017, \apj, 842, 6 
  
\bibitem[Campbell et al.(2013)]{2013Natur.498..198C} Campbell, S.~W., D'Orazi, V., Yong, D., et al.\ 2013, \nat, 498, 198 

\bibitem[Carretta et al.(2013)]{2013ApJ...769...40C} Carretta, E., Gratton, R.~G., Bragaglia, A., et al.\ 2013, \apj, 769, 40 
  
\bibitem[Castelli(2005)]{2005MSAIS...8...25C} Castelli, F.\ 2005, Memorie della Societa Astronomica Italiana Supplementi, 8, 25      

\bibitem[Chantereau et al.(2016)]{2016A&A...592A.111C} Chantereau, W., Charbonnel, C., \& Meynet, G.\ 2016, \aap, 592, A111 
  
\bibitem[Cohen et al.(2011)]{2011ApJ...740...60C} Cohen, J.~G., Huang, W., \& Kirby, E.~N.\ 2011, \apj, 740, 60 
  
\bibitem[Cohen \& Kirby(2012)]{2012ApJ...760...86C} Cohen, J.~G., \& Kirby, E.~N.\ 2012, \apj, 760, 86 
  
\bibitem[D'Antona et al.(2002)]{2002A&A...395...69D} D'Antona, F., Caloi, V., Montalb{\'a}n, J., Ventura, P., \& Gratton, R.\ 2002, \aap, 395, 69 

\bibitem[D'Antona et al.(2016)]{2016MNRAS.458.2122D} D'Antona, F., Vesperini, E., D'Ercole, A., et al.\ 2016, \mnras, 458, 2122 
  
\bibitem[Decressin et al.(2007)]{2007A&A...464.1029D} Decressin, T., Meynet, G., Charbonnel, C., Prantzos, N., \& Ekstr{\"o}m, S.\ 2007, \aap, 464, 1029 

\bibitem[de Mink et al.(2009)]{2009A&A...507L...1D} de Mink, S.~E., Pols, O.~R., Langer, N., \& Izzard, R.~G.\ 2009, \aap, 507, L1 

\bibitem[Denissenkov \& Hartwick(2014)]{2014MNRAS.437L..21D} Denissenkov, P.~A., \& Hartwick, F.~D.~A.\ 2014, \mnras, 437, L21 

\bibitem[Denissenkov et al.(2015)]{2015MNRAS.448.3314D} Denissenkov, P.~A., VandenBerg, D.~A., Hartwick, F.~D.~A., et al.\ 2015, \mnras, 448, 3314 
  
\bibitem[D'Ercole et al.(2008)]{2008MNRAS.391..825D} D'Ercole, A., Vesperini, E., D'Antona, F., McMillan, S.~L.~W., \& Recchi, S.\ 2008, \mnras, 391, 825 
  
\bibitem[D'Ercole et al.(2010)]{2010MNRAS.407..854D} D'Ercole, A., D'Antona, F., Ventura, P., Vesperini, E., \& McMillan, S.~L.~W.\ 2010, \mnras, 407, 854 

\bibitem[Di Criscienzo et al.(2011)]{2011AJ....141...81D} Di Criscienzo, M., Greco, C., Ripepi, V., et al.\ 2011, \aj, 141, 81 

\bibitem[di Criscienzo et al.(2011)]{2011MNRAS.414.3381D} di Criscienzo, M., D'Antona, F., Milone, A.~P., et al.\ 2011, \mnras, 414, 3381 

\bibitem[Di Criscienzo et al.(2015)]{2015MNRAS.446.1469D} Di Criscienzo, M., Tailo, M., Milone, A.~P., et al.\ 2015, \mnras, 446, 1469 
  
\bibitem[Dotter et al.(2008)]{2008ApJS..178...89D} Dotter, A., Chaboyer, B., Jevremovi{\'c}, D., et al.\ 2008, \apjs, 178, 89 

\bibitem[Dotter et al.(2010)]{2010ApJ...708..698D} Dotter, A., Sarajedini, A., Anderson, J., et al.\ 2010, \apj, 708, 698 
  
\bibitem[Gieles et al.(2018)]{2018MNRAS.478.2461G} Gieles, M., Charbonnel, C., Krause, M.~G.~H., et al.\ 2018, \mnras, 478, 2461 

\bibitem[Harris(1996)]{1996AJ....112.1487H} Harris, W.~E.\ 1996, \aj, 112, 1487 

\bibitem[Lagioia et al.(2018)]{2018MNRAS.475.4088L} Lagioia, E.~P., Milone, A.~P., Marino, A.~F., et al.\ 2018, \mnras, 475, 4088 

\bibitem[Lagioia et al.(2018)]{2018arXiv181203401L} Lagioia, E.~P., Milone, A.~P., Marino, A.~F., \& Dotter, A.\ 2018, arXiv:1812.03401 

\bibitem[Lapenna et al.(2015)]{2015ApJ...813...97L} Lapenna, E., Mucciarelli, A., Ferraro, F.~R., et al.\ 2015, \apj, 813, 97 
  
\bibitem[Lee et al.(2013)]{2013ApJ...778L..13L} Lee, Y.-W., Han, S.-I., Joo, S.-J., et al.\ 2013, \apjl, 778, L13 

\bibitem[Karakas et al.(2014)]{2014ApJ...784...32K} Karakas, A.~I., Marino, A.~F., \& Nataf, D.~M.\ 2014, \apj, 784, 32 
  
\bibitem[Krause et al.(2013)]{2013A&A...552A.121K} Krause, M., Charbonnel, C., Decressin, T., Meynet, G., \& Prantzos, N.\ 2013, \aap, 552, A121 
  
\bibitem[Kurucz(2005)]{2005MSAIS...8...14K} Kurucz, R.~L.\ 2005, Memorie della Societa Astronomica Italiana Supplementi, 8, 14
  
\bibitem[Marino et al.(2008)]{2008A&A...490..625M} Marino, A.~F., Villanova, S., Piotto, G., et al.\ 2008, \aap, 490, 625 

\bibitem[Marino et al.(2011)]{2011A&A...532A...8M} Marino, A.~F., Sneden, C., Kraft, R.~P., et al.\ 2011, \aap, 532, A8 
  
\bibitem[Marino et al.(2017)]{2017ApJ...843...66M} Marino, A.~F., Milone, A.~P., Yong, D., et al.\ 2017, \apj, 843, 66 
  
\bibitem[McLaughlin \& van der Marel(2005)]{2005ApJS..161..304M} McLaughlin, D.~E., \& van der Marel, R.~P.\ 2005, \apjs, 161, 304 
  
\bibitem[Milone et al.(2009)]{2009A&A...497..755M} Milone, A.~P., Bedin, L.~R., Piotto, G., \& Anderson, J.\ 2009, \aap, 497, 755 

\bibitem[Milone et al.(2012)]{2012A&A...537A..77M} Milone, A.~P., Piotto, G., Bedin, L.~R., et al.\ 2012, \aap, 537, A77 
  
\bibitem[Milone(2015)]{2015MNRAS.446.1672M} Milone, A.~P.\ 2015, \mnras, 446, 1672 

\bibitem[Milone et al.(2015)]{2015MNRAS.447..927M} Milone, A.~P., Marino, A.~F., Piotto, G., et al.\ 2015, \mnras, 447, 927 
    
\bibitem[Milone et al.(2017)]{2017MNRAS.464.3636M} Milone, A.~P., Piotto, G., Renzini, A., et al.\ 2017, \mnras, 464, 3636 

\bibitem[Milone et al.(2017)]{2017MNRAS.469..800M} Milone, A.~P., Marino, A.~F., Bedin, L.~R., et al.\ 2017, \mnras, 469, 800 
  
\bibitem[Milone et al.(2018)]{2018MNRAS.481.5098M} Milone, A.~P., Marino, A.~F., Renzini, A., et al.\ 2018, \mnras, 481, 5098 

\bibitem[Mucciarelli et al.(2012)]{2012MNRAS.426.2889M} Mucciarelli, A., Bellazzini, M., Ibata, R., et al.\ 2012, \mnras, 426, 2889 
  
\bibitem[Nardiello et al.(2018)]{2018MNRAS.477.2004N} Nardiello, D., Milone, A.~P., Piotto, G., et al.\ 2018, \mnras, 477, 2004 

\bibitem[Prantzos et al.(2017)]{2017A&A...608A..28P} Prantzos, N., Charbonnel, C., \& Iliadis, C.\ 2017, \aap, 608, A28 

\bibitem[Renzini et al.(2015)]{2015MNRAS.454.4197R} Renzini, A., D'Antona, F., Cassisi, S., et al.\ 2015, \mnras, 454, 4197 
  
\bibitem[Sabbi et al.(2016)]{2016ApJS..222...11S} Sabbi, E., Lennon, D.~J., Anderson, J., et al.\ 2016, \apjs, 222, 11
  
\bibitem[Sbordone et al.(2007)]{2007IAUS..239...71S} Sbordone, L., Bonifacio, P., \& Castelli, F.\ 2007, Convection in Astrophysics, 239, 71 

\bibitem[Tailo et al.(2015)]{2015Natur.523..318T} Tailo, M., D'Antona, F., Vesperini, E., et al.\ 2015, \nat, 523, 318 
  
\bibitem[Ventura et al.(2001)]{2001ApJ...550L..65V} Ventura, P., D'Antona, F., Mazzitelli, I., \& Gratton, R.\ 2001, \apjl, 550, L65 

\bibitem[Wang et al.(2016)]{2016A&A...592A..66W} Wang, Y., Primas, F., Charbonnel, C., et al.\ 2016, \aap, 592, A66 

\bibitem[Yong et al.(2008)]{2008ApJ...684.1159Y} Yong, D., Grundahl, F., Johnson, J.~A., \& Asplund, M.\ 2008, \apj, 684, 1159 
  
  
\end{thebibliography}

\end{document}